\documentclass[a4paper]{article}
\usepackage{geometry}
\usepackage{bm}
\usepackage{mathptmx}
\usepackage{graphicx}

\title{Integrating multi-type aberrations from DNA and RNA through dynamic mapping gene space for subtype-specific breast cancer driver discovery}

\begin{document}

\author{Jianing Xi$^1$, Zhen Deng$^2$, Yang Liu$^1$, Qian Wang$^1$, Wen Shi$^{1,*}$\\
\begin{tabular}{l l}
& $^1$ School of Biomedical Engineering, Guangzhou Medical University, Guangzhou, China\\
& $^2$ School of Basic Medical Sciences, Guangzhou Medical University, Guangzhou, China\\
& $^*$ Corresponding author: shiwen@gzhmu.edu.cn
\end{tabular}
}

\date{}

\flushbottom
\maketitle

\begin{abstract}
	Driver event discovery is a crucial demand for breast cancer diagnosis and therapy. Especially, discovering subtype-specificity of drivers can prompt the personalized biomarker discovery and precision treatment of cancer patients. still, most of the existing computational driver discovery studies mainly exploit the information from DNA aberrations and gene interactions. Notably, cancer driver events would occur due to not only DNA aberrations but also RNA alternations, but integrating multi-type aberrations from both DNA and RNA is still a challenging task for breast cancer drivers. On the one hand, the data formats of different aberration types also differ from each other, known as data format incompatibility. One the other hand, different types of aberrations demonstrate distinct patterns across samples, known as aberration type heterogeneity.
	To promote the integrated analysis of subtype-specific breast cancer drivers, we design a "splicing-and-fusing" framework to address the issues of data format incompatibility and aberration type heterogeneity respectively. To overcome the data format incompatibility, the "splicing-step" employs a knowledge graph structure to connect multi-type aberrations from the DNA and RNA data into a unified formation. To tackle the aberration type heterogeneity, the "fusing-step" adopts a dynamic mapping gene space integration approach to represent the multi-type information by vectorized profiles. The experiments also demonstrate the advantages of our approach in both the integration of multi-type aberrations from DNA and RNA and the discovery of subtype-specific breast cancer drivers. In summary, our "splicing-and-fusing" framework with knowledge graph connection and dynamic mapping gene space fusion of multi-type aberrations data from DNA and RNA can successfully discover potential breast cancer drivers with subtype-specificity indication.
\end{abstract}

\thispagestyle{empty}

\section{Introduction}

Breast cancer has become the most prevalent cancer in women worldwide, where 7.8 million women were diagnosed with breast cancer during the past 5 years at the end of 2020 \cite{CA2020}. In cancer cells, driver events in genomics play important roles in tumorigenesis \cite{bailey2018comprehensive}, but the current understanding of breast cancer drivers specially for the poor prognoses triple-negative subtype, is still limited \cite{TCGABRCA}. Consequently, driver event discovery is a crucial demand for breast cancer diagnosis and therapy. Thanks to the unprecedented achievements in DNA sequencing technology \cite{nekrutenko2012next}, the cost of collecting large cohort of cancer genomic data reduces largely, leading to the opportunity for computational discovering cancer driver through the big data of breast cancer samples \cite{ding2014expanding}. In recent researches, numerical studies of computational driver discovery have emerged, but most of the existing studies only focus on the detection of whether the genes to be tested are cancer drivers, lacking of the indication of subtype-specificity \cite{MyBio}. It should be noted that, indicating the subtype-specificity of drivers is an important aspect in precision medicine of cancers \cite{alizadeh2015toward}. Actually, subtype-specificity of drivers can prompt the personalized biomarker discovery and precision treatment of cancer patients \cite{cyll2017tumour}. Therefore, building computational tools for the discovery of subtype-specific drivers is an urge demand for the advancement researches in precision medicine of breast cancer.

Since the most relevant basis of cancer drivers is DNA aberrations in cancer samples, the existing computational driver detection approaches mainly focus on DNA sequencing data, and utilize the information of aberrations like single nucleotide variations \cite{MutSig, OncoClust, WuFangxiang}, copy number alternations \cite{GISTIC, MyTCBB}, and structural variation \cite{SVCell}, etc. In consideration that genes have functional interactions with other genes, some recent approaches for driver discovery also expand the information of genomics from gene mutations to gene interactions \cite{BanjaminReview}. For example, HotNet2 regard the gene interactions as a network with gene nodes \cite{HotNet2}, and propagate the mutations throughout the gene network to integrate the information of mutations and interactions of genes. Subsequently, by restricting propagating step length \cite{MUFFINN} or diffusion scale \cite{ReMIC}, many revisions of network propagation for genomic data integration are also proposed to alleviate the false positives from unrestricted propagation \cite{BanjaminReview}. DawnRank is designed to discover personalized drivers of a single patient by perturbation ranking on the interaction network \cite{DawnRank}. These interaction based driver methods have been widely-accepted as computation tools for driver discovery in recent researches \cite{BanjaminReview}, and they mainly exploit the information from DNA aberrations and gene interactions.

It should be noted that, cancer driver events would occur due to not only DNA aberrations but also RNA alternations \cite{NatureRNA}. For example, PCAWG Transcriptome Core Group has systemically characterized tumor transcriptomes from samples of more than thousands of donors and several cancer genomic databases, and comprehensively analyze the catalogue of cancer-associated RNA alterations of genes \cite{NatureRNA}. They also observe abundance of co-occurrences of RNA and DNA alterations and recurrent RNA alterations in driver genes \cite{NatureRNA}. However, there are few computation tools of cancer driver identification take into account the RNA alternations \cite{NatureRNA}. As demonstrated by PCAWG Transcriptome Core Group, there are also correlations between the DNA and RNA alternations, and these correlations can connect the two types of aberration data to realize the integration of DNA and RNA aberrations \cite{NatureRNA}. Unfortunately, most existing cancer driver detection methods underestimate the information in RNA alternations, only exploit the information from either genomic aberrations or gene interactions. The cancer driver discovery methodology for the integration analysis of the multiple types of aberrations from both DNA and RNA is still lacking.

Nevertheless, for breast cancer driver analysis, integrating multi-type aberrations from both DNA and RNA is a challenging task. On the one hand, the data formats of different aberration types also differ from each other, known as data format incompatibility. Since storing the data into a unified storage is a prerequisite of data integration, the mutations of DNA and altered expressions of RNA are two incompatible formats to each other \cite{TCGABRCA}. One the other hand, different types of aberrations demonstrate distinct patterns across samples, known as aberration type heterogeneity \cite{NatureRNA}. Specifically, different aberration types show mismatched the statistical distributions \cite{LandBreast}, which are difficult to be integrated as one single statistical variable. A widely-used compromise strategy is measuring each type of aberrations into type-specific similarities, and integrating the series of type-specific similarities by weighted summation \cite{MaXiaoke, BaDan1, BaDan2} or regularization \cite{MyPan}. But there is still a drawback for similarity-based integrations since they would cause the degradation of multi-type aberrations, leading to only a scalar result with deficiency of the informative multi-dimension aberrations \cite{MyPan}. Therefore, how to integrate aberrations from both DNA and RNA into a unified format with the informative multi-type aberration heterogeneity preserving, is still a bottleneck for driver discovery from multi-type aberrations.

To promote the integrated analysis of subtype-specific breast cancer drivers, we design a "splicing-and-fusing" framework to address the issues of data format incompatibility and aberration type heterogeneity respectively. In the "splicing-step" of our framework, we firstly adopt knowledge graph structure to reformat the DNA and RNA data into a unified formation of multi-type aberrations \cite{ProIEEE}, which can successfully overcome the data format incompatibility by connecting the information from different sources as a series of triplets of facts. Furthermore, we also propose a dynamic mapping \cite{TransD} gene space integration approach in the "fusing-step". In consideration of the aberration type heterogeneity, this gene space approach can represent the multi-type information into a vectorized profile, instead of scalar representation \cite{ReviewKG}. To evaluate the efficiency of our integration analysis approach for subtype-specific drivers on breast cancer data \cite{TCGABRCA}, we conduct experiments of comparison study to assess the cancer driver discovery performance, and ablation studies on both data and methods to assess the effects of multi-type information and dynamic mapping strategy. We also employ experiment of data visualization for subtype-specificities indication of the discovered potential drivers, and further unscrambling gene functions by enrichment analysis experiment. The experimental results indicate the superiority of our approach in the discovery of subtype-specific breast cancer driver, and the advantage of integrating multi-type aberrations from both DNA and RNA for cancer drivers.

\section{Materials and Methods}

\subsection{Data Acquisition of Breast Cancer Samples}

The data of DNA and RNA aberrations of breast cancer samples are collected from The Cancer Genome Atlas (TCGA) project, a well-curated database including DNA sequencing data and RNA expression data of cancer samples \cite{TCGABRCA}. The cancer samples with both DNA and RNA data available are selected as our integration analysis, where the total volume of 523 breast cancer samples. To simply the preprocessing of the TCGA data, we adopt a pre-compiled source, of the TCGA breast cancer data from the UCSC Xena platform for cancer genomics data \cite{UCSC}. The UCSC Xena platform provides the occurrence of gene aberrations in DNA of each sample and the abundance of gene expressions in RNA of each sample \cite{UCSC}. Specifically, there are several different types the occurrence of gene aberrations in DNA, including 3' Flank, 3' UTR, 5' Flank, 5' UTR, frame shift del, frame shift ins, IGR, in frame del, in frame ins, intron, missense mutation, nonsense mutation, nonstop mutation, silent, splice region, splice site, and translation start site. At the same time, the abundance of gene expressions in RNA is also utilized in mining differentially expressed RNAs of genes \cite{XingMing}, including over-expressions and under-expressions. These differentially expressed RNAs of genes are regarded as RNA alternations of genes. Accordingly, the aberration data in DNA and RNA are aligned to their corresponding breast cancer samples, and the samples can play the roles as the anchor between the two distinct types of aberrations. The collected aberrations from both DNA and RNA data of each sample are the basis of data integration.

\begin{figure}[h!]
	\begin{center}
		\includegraphics[width=\linewidth]{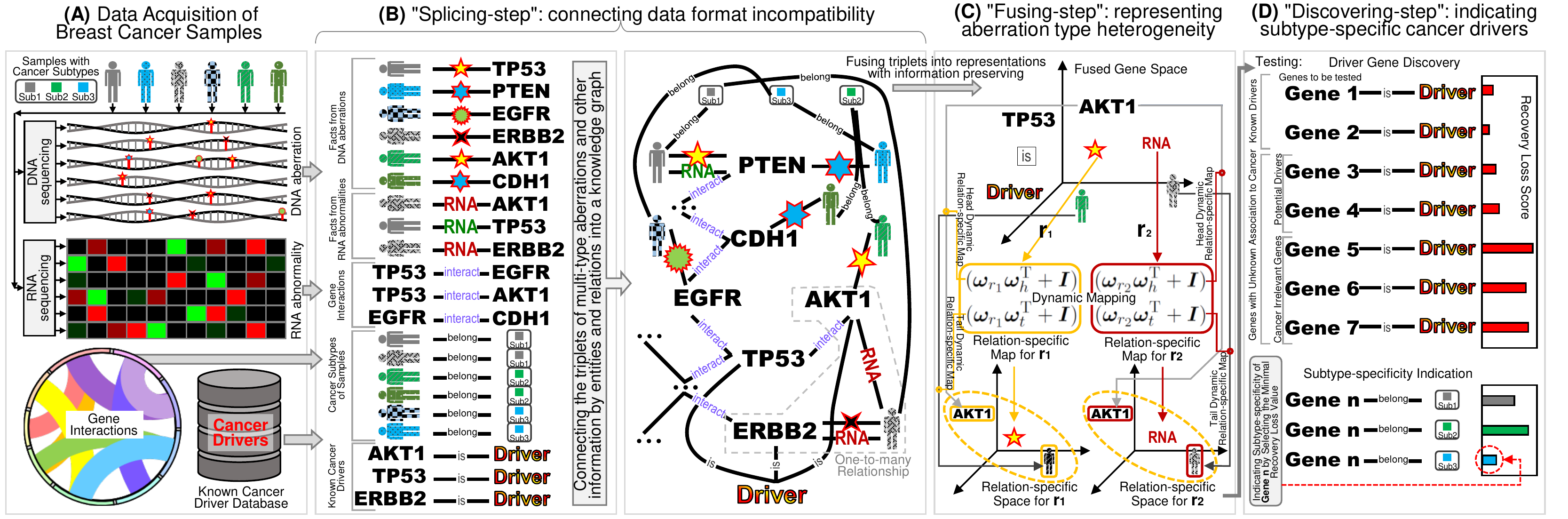}
	\end{center}
	\caption{The schematic diagram of our "splicing-and-fusing" framework for subtype-specific cancer driver discovery. (A) Data acquisitions of DNA and RNA aberrations data from breast cancer samples, gene interactions, and known benchmarking drivers. (B) The "splicing-step" connecting data format incompatibility via constructing the knowledge graph. (C) The "fusing-step" representing aberration type heterogeneity via dynamic mapping relation-specific gene space. (D) The "discovering-step" indicating subtype-specific cancer drivers via recovery loss scores.}\label{fig:schematic}
\end{figure}

In addition to aberration data in DNA and RNA, we can also introduce side information to prompt the subtype-specific breast cancer driver discovery task in our integration analysis. For subtype-specificity indication task in our integration, the TCGA database also provides the subtype annotations of the breast cancer samples, including five widely-accepted intrinsic molecular subtypes: Luminal A, Luminal B, HER2-enriched, Triple-negative, and Normal-like subtypes \cite{TCGABRCA}. The subtype annotations of the breast cancer samples can also be obtained from the UCSC Xena platform \cite{UCSC}. Here the subtype annotations can play the role as side information for subtype-specificity indication of the investigated genes. Furthermore, in consideration that the gene interactions also play important roles in tumorigenesis \cite{BanjaminReview}, accordingly we also incorporate the gene interaction data into our integration analysis as network format data, regarding the genes as network nodes and the interactions as network edges \cite{HotNet2}. The gene interaction information is collected from two classical database, STRING \cite{STRING} and iRefIndex \cite{iRefIndex} , facilitating the customizing integration with different versions of curated interactions. The annotations of genes being known cancer drivers are also introduced as the supervision information to support the cancer driver discovery in the integration analysis. In summary, we include gene aberrations from DNA, gene aberrations from DNA, subtype annotations of samples, gene interactions, and known cancer driver annotations of genes as the data used in the integration analysis (\textbf{Figure~\ref{fig:schematic}A}).

\subsection{Overview of "Splicing-and-fusing" Framework}

Since the RNA alternation and DNA mutations are distinct types of aberrations of genes, their integration faces the challenges of data format incompatibility and aberration type heterogeneity. Therefore, we design a "splicing-and-fusing" framework to address the two challenges step by step. After the data collection (\textbf{Figure \ref{fig:schematic}A}), we firstly feed the data into "splicing-step". To overcome data format incompatibility issue, in the "splicing-step" the multi-type data from DNA, RNA and other sources are redescribed as a series of facts of triplets, and the data of facts are spliced together by connections across the triplets (\textbf{Figure~\ref{fig:schematic}B}). Therefore, the "splicing-step" can successfully address the problem of data format incompatibility and joint the multi-type data into a flexible and unified format as knowledge graph. The next step of our "splicing-and-fusing" framework is the "fusing-step" (\textbf{Figure~\ref{fig:schematic}}C). In this step, we introduce the data representation of knowledge graph and adopt a dynamic mapping gene space fusing strategy. The idea of dynamic mapping can prompt the relation-specific resolution of data fusion for many-sided representations, and the gene space representations can describe the multi-type information into vectorized profiles. Compared with scalar score, our vectorized gene space can offer an informative many-sided representation to describe the aberration type heterogeneity. After the dynamic mapping gene space representations are obtained, we further apply the indication of subtype-specific cancer drivers in "discovering-step" (\textbf{Figure~\ref{fig:schematic}D}). Here we use the recovery loss score learnt during "fusing-step" to infer whether the investigated genes are a driver or not and indicate which subtype the drivers would belong to. Finally, we can obtain the breast cancer drivers from the investigated genes as well as their subtype-specificities.


\subsection{"Splicing-step": connecting data format incompatibility}

To address the data format incompatibility of multi-type aberrations, we convert the data of each type as a series of itemized facts \cite{ProIEEE}, and splice these facts as a knowledge graph. Specifically, for both DNA and RNA, we can record a fact as a triplet (sample $P$, aberration type $T$, gene $G$) if gene $G$ occurs with a $G$ type aberration in sample $T$ (\textbf{Figure~\ref{fig:schematic}B}). Here a triplet fact includes three elements: subject, predicate, and object. Here the subject and object are the head $h$ and tail $t$ of the triplet fact respectively, and the predicate is the relation $r$ between the head $h$ and tail $t$. When we list all the triplet facts of aberrations from both DNA and RNA, every type of aberrations in their related genes of all the samples are fully covered in these facts. Here the subjects include all samples and the objects include all aberrated genes, which are regarded as entities in the triplets. The aberration types between the samples and genes are regarded as relations in the triplets. When we collect all the entities as graph nodes and relation as graph edges, the connections between the entities and relations can be formed as a graph structure with these different types of nodes and edges as a knowledge graph \cite{ProIEEE}. Since the edges in the knowledge graph can preserve different types of relations, this connection structure can overcome the incompatibility of different data formats of aberrations from DNA and RNA.

Since the information related to breast cancer drivers is far beyond the aberrations of DNA and RNA, such as the interactions between genes and the subtype annotations of breast cancer samples \cite{BanjaminReview}, we also include the side information into the connection structure of knowledge graph. Specifically, the knowledge graph structure can also incorporate gene interactions by adding the triplet facts of gene interaction relationships. Here we can format a triplet (gene $G_1$, interaction $I$, gene $G_2$) if there exists an interaction between gene $G_1$ and $G_2$ (\textbf{Figure \ref{fig:schematic}B}). Then the facts of gene interactions can be easily included in the knowledge graph by only listing additional triplets related to interaction. Note that many genes may have several synonym names \cite{DAVID}, we can also list the synonym facts (gene $G_1$, synonym $S$, gene $G_2$) and feed these triplets into the knowledge graph. When all the triplet facts are listed together, we can obtain the entity set of all the heads (subjects) and tails (objects), and the relation set of all the predicates (multi-type aberrations between samples and genes, and interactions/synonyms between genes). Another type of side information is the triplets related to subtypes of breast cancer \cite{TCGABRCA}. These subtype annotations of samples can be formed as triplets (sample $P$, belong to $B$, subtype $S$), when the investigated breast cancer sample $P$ (subject) belongs to (predicate) the certain subtype $S$ (object) (\textbf{Figure~\ref{fig:schematic}B}). The inclusion of subtype annotations plays the key role in the subtype-specificity indication in the next subsection.

Finally, we also add the facts of the existing known cancer drivers \cite{CGC} by triplets (gene $G$, is $Is$, driver $D$) (\textbf{Figure~\ref{fig:schematic}B}). These facts also play the roles of supervision information in the constructed knowledge graph. Note that it is still unknown whether many aberrant genes are associated with cancer or not, but these genes are included in the triplets as entities due to their participation in genomic aberrations or gene interactions \cite{CGC}. Accordingly, the collected triplets of the facts that some genes are known cancer drivers can only cover a subset of driver genes, and some other genes in the triplets are potential drivers to be discovered. Here a collected triplet with cancer driver annotation (gene $G$, is $Is$, driver $D$) can indicate that the gene $G$ in this triplet is an experimental validated known cancer driver, but the absence of a driver annotation triplet of a gene cannot indicate its irrelevance to cancer driver \cite{ReviewKG, ConfT}. Actually, the absence of driver annotation can only tell that we do not know whether the gene is driver or not. Through the collections of triplets aforementioned, when we regard the same entities and relations across these triplets as graph nodes, we can connect the triplets of facts as graph edges (\textbf{Figure~\ref{fig:schematic}B}). The connection result can form an integrated knowledge graph including a series of triplet facts of multiple types of aberrations from both DNA and RNA. Since this step contains the operation of splicing together different triplet facts into an integrated knowledge graph, we denote this multi-type aberration data connection step as "splicing-step", and this knowledge graph based splicing-step via can successfully overcome the data format incompatibility problem.

\subsection{"Fusing-step": representing aberration type heterogeneity}

To preserve the aberration type heterogeneity in the integrated representation, we further propose a dynamic mapping gene space integration approach, known as the "fusing-step". Unlike the previous scalar based methods such as statistical variation approaches \cite{MutSig, OncoClust} or similarity based integration approaches \cite{HotNet2, MUFFINN}, we select a vectorized profile containing multiple dimensions to represent the information of genes with multi-type aberrations. In the vectorized profiles of gene representations, each gene denotes a multi-dimension vector that can contain various trends across different samples in a mapped multi-dimension gene space. Fortunately, there are several advanced methods that can support the multi-dimension vector representations of knowledge graph, throwing lights on the learnt space representations of multi-type aberrations \cite{ReviewKG}. Inspired by the representation learning advancements, we can fuse the knowledge graph containing multiple types of aberrations into an integrated gene space with the graph structure preserved, known as knowledge graph embedding \cite{ReviewKG}. Thus, in this paper, we propose a dynamic mapping strategy to embed the knowledge graph into integrated gene space for fusing the DNA and RNA aberrations.

Technically, the learnt fused representations are expected to successfully recover the raw triplets in the multi-type aberration knowledge graph, i.e., the fusing strategy should be information preserving \cite{ProIEEE}. At the same time, the learnt fused representation of the knowledge graph should also be in a low-dimensional space \cite{ReviewKG}. This is due to the hypothesis that the lower the learnt gene space dimension, the more concise the computational discovery would be \cite{ReviewKG}. Consequently, fusing the aberration information can be formulated as a task that using a series of vectors in a learnt space to approximately recover the known triplets in knowledge graph \cite{ReviewKG}. Specifically, a common strategy for triplet information fusion is to represent the relationships of entities and relations in the triplets through translations operating on the low-dimensional representations \cite{ReviewKG, TransE}. By optimizing the translations between the head and tail on a certain relation through a distance metric score as $Dis^{r}(h, t)$, we can obtain a series of vectors for entities and relations fusing the information in a representation space of genes.

For putting the idea of information fusion by triplet translations operating on the low-dimensional representations into practice, a straightforward but effective approach for learning a translation representation space of fused triplets is to minimizing the residual of head and tail vectors by summation \cite{TransE}. Detailly, the specific translation for a certain triplet $(h, r, t)$ can be modeled as a vector summation in the representation space as $\bm{v}_h + \bm{v}_r \approx \bm{v}_t$. This approximation is expected to induce the information preserving of the learnt representation vectors (also known as embedding vectors) of the entities and relations in the knowledge graph \cite{TransE}. A common choice is to minimize the L1-norm on the vector residual $\bm{v}_h + \bm{v}_r - \bm{v}_t$, formed as the distance between the head entity $h$ and tail entity $t$ on a certain relation $r$ in the fused triplet $(h, t, r)$:

\begin{equation}
	\mathrm{Dis}^{(r)}(\bm{v}_h, \bm{v}_t) = \left\| \bm{v}_h + \bm{v}_r - \bm{v}_t \right\|_1,
	\label{eq:01}
\end{equation}

\noindent where the subscript $1$ denotes that the metric is set to L1-norm in consideration of its sparse penalty property in the role of distance \cite{TransE}. The summation based fusing method, also known as TransE \cite{TransE}, can provide the representations as translations in the fused vector space.

Despite the efficiency of the simple translational fusing model \cite{TransE}, a shortcoming for the summation based translational fusing approach is that this method cannot distinguish the representations for one-to-many triplets, thanks to the uniqueness of summation \cite{TransH}. However, in a breast cancer sample, different types of aberrations are likely to occur in one certain gene at the same time, and this phenomenon shows that one-to-many triplets are inevitable in multi-type aberration knowledge graph data \cite{TransH}. Consequently, the representations of entities in gene space should be many-sided when confronting multi-type aberrations as multiple relations. In consideration that different entities in one-to-many triplets are usually associated with distinct relations, a plausible solution is projecting the representation vectors to different relation-specific hyperplanes (also known as mapping on planes), so that the entity vector can be represented as many-sided in fused gene space \cite{TransH}. Specifically, when an entity in one-to-many triplet confronts various relations, the information preserving of summation approximation can be revised as a summation of the projected vectors on relation-specific hyperplanes rather than a simple vector summation in the gene space:

\begin{equation}
	\mathrm{Dis}^{(r)}( \bm{v}_h^{\mathrm{proj}}, \bm{v}_t^{\mathrm{proj}} ) = \left\| (\bm{v}_h - \bm{\omega}^\mathrm{T}_r \bm{v}_h \bm{\omega}_r) + \bm{v}_r – (\bm{v}_t - \bm{\omega}^\mathrm{T}_r \bm{v}_t \bm{\omega}_r)\right\|_1,
	\label{eq:02}
\end{equation}

\noindent where the vector $\bm{\omega}_r$ is the normal vector of a relation-specific hyperplane, and $\bm{v}_h^{\mathrm{proj}} = (\bm{v}_h - \bm{\omega}^\mathrm{T}_r \bm{v}_h \bm{\omega}_r)$ and $\bm{v}_t^{\mathrm{proj}} = (\bm{v}_t - \bm{\omega}^\mathrm{T}_r \bm{v}_t \bm{\omega}_r)$ are the projection vectors of representations of head and tail entities in the fused gene space. This translating on hyperplanes fusing strategy, also known as transH \cite{TransH}, is a breakthrough to tackle multi-relation data. Through the relation-specific hyperplane projection in gene space, we can circumvent the multiple relations in one-to-many triplets in the learnt fused information.

To further enhance the information preserving of data fusing for multi-type aberration knowledge graph with one-to-many or many-to-many triplets, we can also introduce a relation-specific space instead of a hyperplane \cite{TransR, TransD}. In comparison with hyperplanes, relation-specific spaces can cover more fine-grained connotations for entities when their representation vectors are mapped into various situations in one-to-many or many-to-many triplets \cite{TransR}. An intuitive solution to construct a relation-specific space is to build a static map from the fused representation gene space to the relation-specific space \cite{ReviewKG}. Specifically, the coefficients of the static map can be described as a series of elements in a relation-specific matrix $\bm{M}_r$, and the linear transformation can be constructed as matrix product such as $\bm{v}_h^{ StaticMap } = \bm{M}_r \bm{v}_h$ and $\bm{v}_t^{ StaticMap } = \bm{M}_r \bm{v}_t$ for heads and tails respectively \cite{TransR}. Under the idea of relation-specific spaces, the information preserving of approximation is the distances between the head and tail entities on the relation-specific spaces are formulated as:

\begin{equation}
	\mathrm{Dis}^{(r)}( \bm{v}_h^{ \mathrm{ StaticMap} }, \bm{v}_t^{ \mathrm{StaticMap }} ) = \left\| (\bm{M}_r \bm{v}_h) + \bm{v}_r –(\bm{M}_r \bm{v}_t) \right\|_1.
	\label{eq:03}
\end{equation}

This idea of information fusion by translation in the corresponding relation space is also known as TransR \cite{TransR}. However, when the number of relation types is large, the degree of freedom of the parameters in relation space map in the relation-specific matrix also grows rapidly, causing the underdetermined problem and overfitting risk \cite{KDD}.

To reach a higher relation-specific resolution data fusion and a dynamic many-sided representation of the head and tail of triplet data, we further introduce the idea of dynamic mapping matrix \cite{TransD} into the relation-specific space fusing multi-type aberration knowledge graph. Rather than using a relation-specific matrix with full degree-of-freedom static parameters, we choose a compression scheme \cite{TransD} to build a relation-specific space with a mapping matrix that are dynamic for the different entities and relations (\textbf{Figure \ref{fig:schematic}C}). For each triplet in multi-type aberration knowledge graph, the parameter matrix of the dynamic mapping is computed by projection vectors of both entity and relation. To ensure fine-grained fusion representations in the multi-type aberration triplets \cite{TransD}, the projections are also dynamic for the entities between heads and tails as $\bm{M}_r^h = \bm{\omega}_r\bm{\omega}_h^\mathrm{T} + \bm{I}$ and $\bm{M}_r^t = \bm{\omega}_r\bm{\omega}_t^\mathrm{T} + \bm{I}$ respectively. Accordingly, the information preserving of approximation for dynamic mapping can simultaneously incorporate the diversity of entities and relations (\textbf{Figure \ref{fig:schematic}C}):

\begin{equation}
	\mathrm{Dis}^{(r)}( \bm{v}_h^{ \mathrm{DynMap} }, \bm{v}_t^{ \mathrm{DynMap} } ) = \left\| [(\bm{\omega}_r\bm{\omega}_h^\mathrm{T} + \bm{I}) \bm{v}_h] + \bm{v}_r – [(\bm{\omega}_r\bm{\omega}_t^\mathrm{T} + \bm{I}) \bm{v}_t] \right\|_1,
	\label{eq:04}
\end{equation}

\noindent where $\bm{v}_h^{ DynMap } = \bm{M}_r^h \bm{v}_h$ and $\bm{v}_t^{ DynMap } = \bm{M}_r^t \bm{v}_t$. To fuse the information in the triplets of the knowledge graph, we can perform the Adam optimizer \cite{Adam} on the joint loss function for recovering all the existing triplets of facts:

\begin{equation}
	\min_{(\bm{v}_h, \bm{v}_r, \bm{v}_t, \bm{\omega}_h, \bm{\omega}_r, \bm{\omega}_h)} \sum_{(h, r, t) \in \mathcal{G}}\mathrm{Dis}^{(r)}( \bm{v}_h^{ \mathrm{DynMap} }, \bm{v}_t^{ \mathrm{DynMap} } ),
	\label{eq:05}
\end{equation}

\noindent where the symbol $\mathcal{G}$ denotes the set containing all the triplets of facts in the integrated knowledge graph. Finally, after the information preserving representations are learnt via optimizing the aforementioned function, the one-to-many and many-to-many triplets with multiple relations can dynamically represent the multi-type aberration knowledge graph through the dynamic mapping gene space (\textbf{Figure \ref{fig:schematic}C}).

\subsection{"Discovering-step": indicating subtype-specific cancer drivers}

It should be noted that the motivation of integrating multi-type aberrations from DNA and RNA is not for data integration itself, but for the clinical application of subtype-specific breast cancer driver discovery. Due to the fact that all the genes in genomics are contained in the aberration knowledge graph in the "splicing-step" of the integration, both the known cancer driver genes and the genes with unknown association with cancer are all included in the graph \cite{ConfT}. Consequently, the task of cancer driver discovery can be equivalent to disclosing the potential driver genes from the genes with unknown association to cancer, since there are also cancer irrelevant genes included. Although the genes to be tested are already included in the multi-type aberration knowledge graph, the entities of the potential driver genes do not have known associations with the cancer driver entity, and hence the knowledge graph in established "splicing-step" cannot discover the potential cancer drivers directly. Fortunately, in the "fusing-step", the fused representations in the gene space of multi-type aberration knowledge graph can be applied in cancer driver discovery, thanks to the link prediction application of graph embedding \cite{LinkPred, KDD}. In the "fusing-step", the fused representations in the gene space not only can recover the known triplets of facts in the multi-type aberration knowledge graph, but also can discover the potential driver genes of breast cancer.

For the cancer driver discovery by the dynamic mapping gene space, the reason for utilizing the fused information of the gene entities is that their representations in dynamic mapping gene space can recover or predict their relations with driver entity. Since the fusing process not only optimize the recovery loss of the existing known triplets but not the unknown triplets, the potential relations between the genes and the entity of cancer driver are not restricted to be negative \cite{ReviewKG, ConfT}, but also feasible to be positive if the genes are potential drivers \cite{KTen, ConfT}. Consequently, by computing the recovery loss function on the genes to be tested \cite{LinkPred}, we can obtain the loss score $Dis^{r}(h, t)$ of these genes across the triplets with driver entity (\textbf{Figure \ref{fig:schematic}D}), where the tested triplet $(h, r, t)$ is $(h, r, t) = $(gene to be tested $G_\mathrm{test}$, is $Is$, driver $D$). If the loss score $Dis^{Is}( G_\mathrm{test}, D)$ is smaller, then the tested gene $G_\mathrm{test}$ has more potential to be a cancer driver. The basis that the recovery loss score has the capability of reflecting cancer drivers is the inclusion of known drivers as supervision information in the knowledge graph data \cite{WuFangxiang}. Thus, the potential driver genes tend to have small values of the recovery loss scores $\mathrm{Dis}^{Is}( G_\mathrm{test}, D)$ in triplets (gene to be tested $G_\mathrm{test}$, is $Is$, driver $D$). The recovery loss scores are inducted to yield small values for all the triplets of known drivers in the knowledge graph data, and therefore the scores of potential cancer driver genes are also expected to be small for unobserved triplets including the genes to be tested \cite{WuFangxiang}. Consequently, we can apply the fused data representation from multi-type aberration knowledge graph on the discovery of potential breast cancer drivers.

Since only inferring whether a gene to be tested is a potential driver is not enough for the indication of subtype-specific drivers for breast cancer, therefore we should also indicate the subtype-specificity of the discovered driver genes \cite{MyBio}. Note that there are no triplets of facts describing the indication relationships of driver genes to cancer subtypes as the form (gene $G$, belong to $B$, subtype $S$), But we can borrow the known information from the subtype belongingness of samples with the triplet form (sample $S$, belong to $B$, subtype $S$). Here the fused representations in gene space are expected to reflect the subtype information due to the inclusion of subtype related triplets in knowledge graph. Like the idea of cancer driver discovery above, we can also adopt the recovery loss scores to indicate the subtype-specificity of the investigated gene. For a potential driver gene $G$ discovered by the fused representations, we further utilize the recover loss score function $\mathrm{Dis}^{\cdot}(\cdot, \cdot)$ on the unknown triplet (gene $G$, belong to $B$, subtype $S$). Accordingly, we can yield the scores of the investigated gene $G$ across all the subtypes $S_1, S_2, \ldots, S_K$, as $\mathrm{Dis}^{B}(G, S_1), \mathrm{Dis}^{B}(G, S_2), \ldots, \mathrm{Dis}^{B}(G, S_K)$. Since a small score indicates a better change of the unknown triplet, we can compare the scores across different subtypes, and assign subtype for the smallest score to the investigated gene as its subtype-specificity (\textbf{Figure \ref{fig:schematic}D}). Based on the recovery loss score assignment, we can achieve the subtype-specificity indication of cancer drivers through the fused representation gene space.

\section{Experimental Results}
\subsection{Experiment setup}
\subsubsection{Experiment design} 
To examine the effectiveness for subtype-specific cancer driver discovery of our multi-type aberration integration approach, we conduct several experiments to analyze our approach from multiple perspectives: 1) comparing our approach with widely-used existing cancer driver approaches, 2) testing the effect of integrating multi-type aberrations by data ablation, 3) applying ablation study for gene space by removing dynamic mapping strategies, 4) indicating subtype-specificities of genes by data visualization, and 5) unscrambling gene functions by enrichment analysis. The first three experiments are conducted for examining the cancer driver discovery power by the gold standards of known benchmarking driver genes, in order to investigate the advancement of our approach, the effect of different type information, and the effect of data fusing strategies. Specifically, we compare our approach with existing interaction based driver methods including HotNet2 \cite{HotNet2}, DawnRank \cite{DawnRank}, two versions of MUFFINN (DNsum version and DNmax version respectively) \cite{MUFFINN}, and our previous method DriverSub (including settings of k=3 and k=4) \cite{MyBio}. The fourth experiment is performed to visualizingly demonstrate the indication of subtype-specificities of drivers. The fifth experiment is applied to unscramble the gene functions of discovered drivers.

\subsubsection{Evaluation Metrics}
In the evaluation of cancer driver discovery, there are gold standards of known benchmarking driver genes, and consequently we easily examine whether a discovered gene is truly a cancer driver. Note that during the "splicing-step" of our approach, the information of the known cancer driver is a part of triplets of the facts include in the integrated knowledge graph as the supervision information. Thus, we employ the idea of cross validation for cancer driver discovery performance evaluation \cite{WuFangxiang}. Specifically, we evenly split the known cancer driver triplets into five folds, and select one of the five folds of triplets out of the integrated knowledge graph as testing set. The remaining four folds of driver triplets are included in the integrated data as training set. We further selected the other folds as testing sets and keep the remains for training, and all the five folds can serve as five-fold cross validation. Since the comparison including both supervised methods and unsupervised methods, the outputs of genes to be tested in each fold are concatenated together so that the concatenated outputs of every gene are free from supervision information \cite{KDD}. Accordingly, the outputs of genes are evaluated by Receiver Operating Characteristic (ROC) curves, where the x-axis of the curve is the False Positive Rate (FPR) which is the fraction of discover genes in the non-benchmarking drivers \cite{CGC}, and the y-axis of the curve is the True Positive Rate (TPR) which is the fraction of discover genes in the benchmarking drivers. As for visualization for subtype-specificity and enrichment analysis for unscrambling functions, there are no quantitative evaluation metrics, and we focus on the illustration of the results in these two experiments. 

\subsection{Comparison study for driver discovery}

\begin{figure}[h!]
	\begin{center}
		\includegraphics[width=\linewidth]{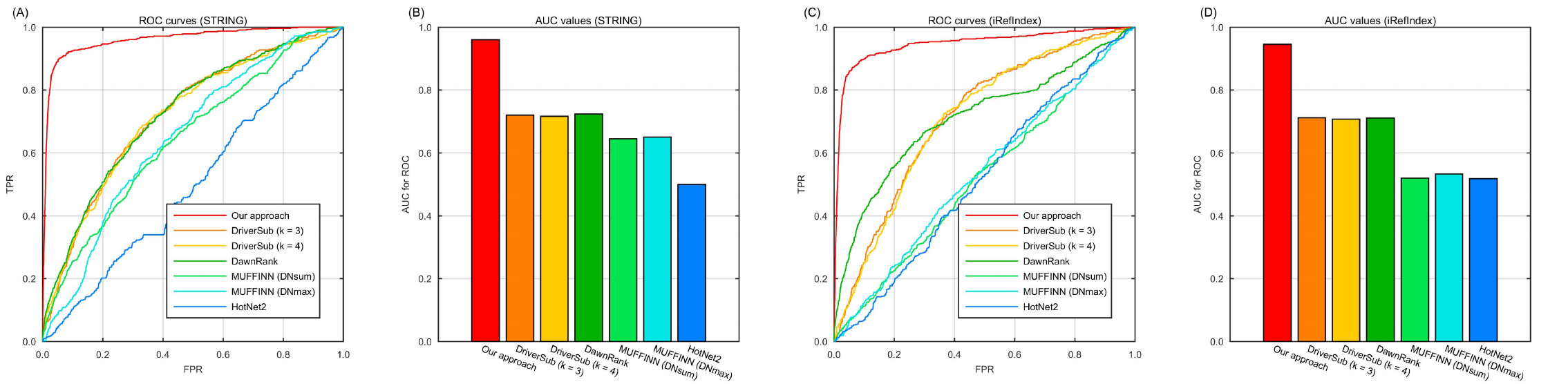}
	\end{center}
	\caption{The ROC curves of our approach against those of the existing methods on breast cancer data and gene interactions. (A) ROC curves of the competing methods with interaction source of STRING. (B) AUC values of the competing methods with interaction source of STRING. (C) ROC curves of the competing methods with interaction source of iRefIndex. (D) AUC values of the competing methods with interaction source of iRefIndex.  }\label{fig:comp}
\end{figure}

To demonstrate the cancer driver discovery performance of our dynamic mapping gene space integrated multi-type aberration information, we adopt our approach on the TCGA breast cancer data \cite{TCGABRCA} by constructing a knowledge graph and fusing the multi-type information to yield the recovery loss scores of the investigated genes. We also compare the widely-used existing interaction based driver approach HotNet2 \cite{HotNet2}, DawnRank \cite{DawnRank}, MUFFINN \cite{MUFFINN}, and our previous study DriverSub \cite{MyBio} on the breast cancer dataset. Since we have two choices of gene interaction resources, STRING \cite{STRING} and iRefIndex \cite{iRefIndex}, here we demonstrate the performance of these competing methods under the inclusion of the two interaction resources separately. Through the ROC curve evaluation, we can observe that our approach achieves the best discovery performance of those of the competing methods. Specifically, when using the interaction resource of STRING, we can observe that DriverSub and DawnRank achieve better performance than those of HotNet2 and the two versions MUFFINN (\textbf{Figure \ref{fig:comp}A}). Accordingly, the AUC values of DriverSub (k=3), DriverSub (k=4), and DawnRank are located in the range from 70.0\% to 80.0\% (\textbf{Figure \ref{fig:comp}B}). In comparison, the curve of our approach is closest to the top-left corner (\textbf{Figure \ref{fig:comp}A}), yielding an AUC value of 96.1\%. As for the case of interaction resource of iRefIndex, we can observe a similar phenomenon that the curve of our approach fully covers the areas of the curves of the other competing methods (\textbf{Figure \ref{fig:comp}C-D}). A plausible explanation of the phenomena is that in comparison with the other existing methods, our approach integrates multi-types of information beyond the information of only DNA aberrations and gene interactions. Generally, our approach demonstrates a favorable performance on the task of cancer driver discovery.

\subsection{Data ablation for multi-type aberrations}

\begin{figure}[h!]
	\begin{center}
		\includegraphics[width=\linewidth]{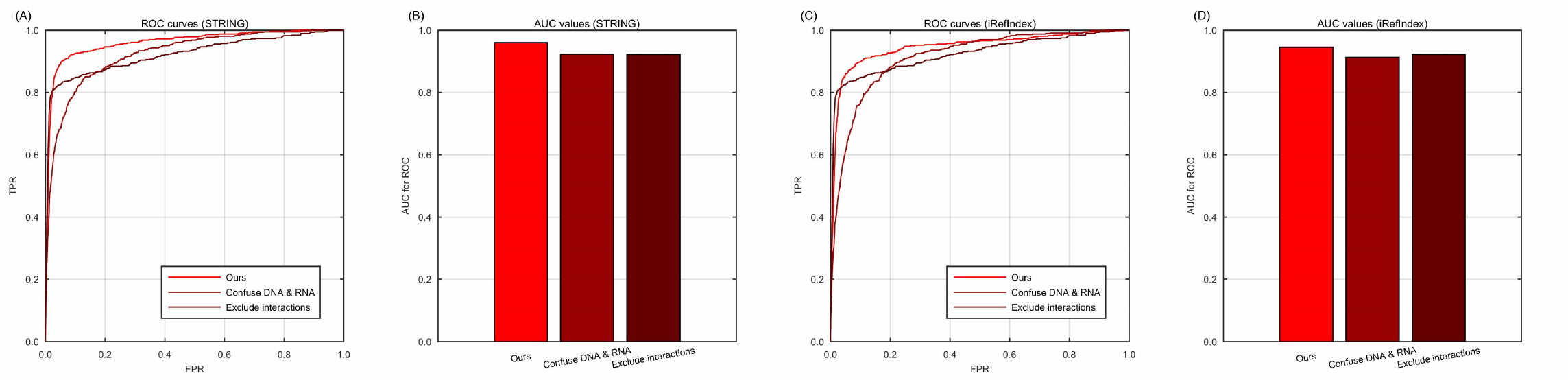}
	\end{center}
	\caption{The ROC curves of our approach for data ablation on breast cancer data and gene interactions. (A) ROC curves for data ablation  with interaction source of STRING. (B) AUC values for data ablation with interaction source of STRING. (C) ROC curves for data ablation with interaction source of iRefIndex. (D) AUC values for data ablation with interaction source of iRefIndex.}\label{fig:ablationData}
\end{figure}

To further investigate the effects of information integrations from multi-type data, we further conduct the data ablation on the multi-type data. Here data ablation is a strategy for observing the performance effects when the data are changed \cite{mousavi2020ai}. When we confuse the multi-type aberrations from DNA and RNA as one single type, i.e., only considering whether the genes are with aberrations but excluding the difference of aberration types between DNA and RNA. In the knowledge graph, the relations of different aberration types are confused as only one type of aberration, but the other types of relations such as interaction or subtype belongingness are not confused. As shown in \textbf{Figure \ref{fig:ablationData}A} for STRING and \textbf{Figure \ref{fig:ablationData}C} for iRefIndex, the curve of confusing multi-types is lower than that of the full version data of our approach, i.e., no aberration type confusing. These phenomena indicate that the difference between multi-type aberrations are informative for the driver discovery of our integration approach. Another way to change the integrated data is to exclude the information of gene interactions from the knowledge graph. By this way, we can observe the effects of performance from the interaction information. Through \textbf{Figure \ref{fig:ablationData}A} and \textbf{\ref{fig:ablationData}C}, we can find that the performance decrease by excluding interactions are comparable with the decrease of confusing aberration types. Here the two curves in \textbf{Figure \ref{fig:ablationData}A} and \textbf{\ref{fig:ablationData}C} are exactly the same due to their knowledge graphs are also exactly the same when the interactions are removed. The AUC values of the curves of the cases in data ablation are also provided in \textbf{Figure \ref{fig:ablationData}}B and D for interaction resource with STRING or iRefIndex respectively.

\subsection{Ablation study for gene space mapping}

\begin{figure}[h!]
	\begin{center}
		\includegraphics[width=\linewidth]{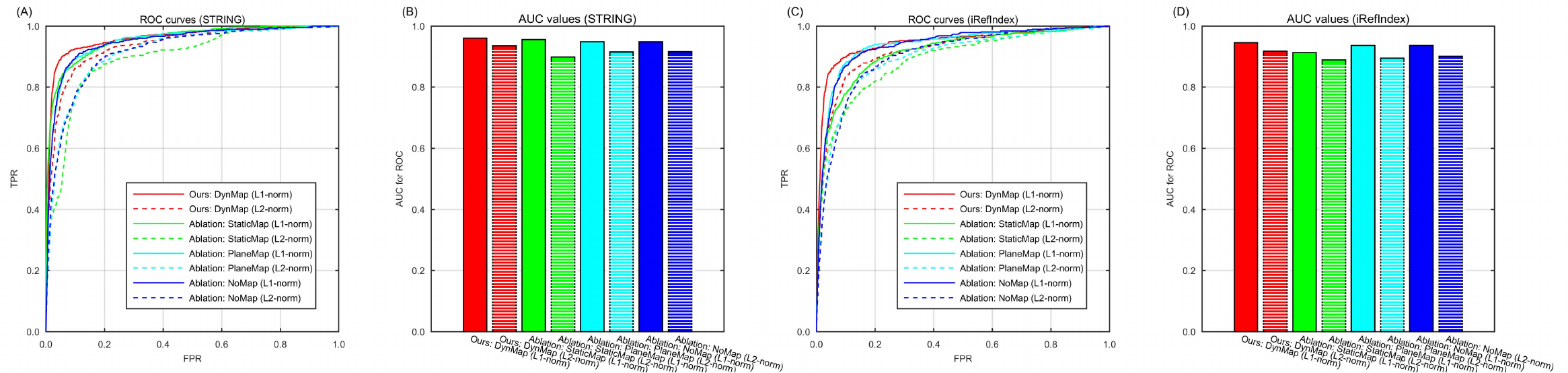}
	\end{center}
	\caption{The ROC curves of our approach to ablation study in methodology on breast cancer data and gene interactions. (A) ROC curves for ablation study in methodology with interaction source of STRING. (B) AUC values for ablation study in methodology with interaction source of STRING. (C) ROC curves for ablation study in methodology with interaction source of iRefIndex. (D) AUC values for ablation study in methodology with interaction source of iRefIndex.}\label{fig:ablationMeth}
\end{figure}

In addition to data ablation that excluding part of the multi-type data, we also conduct the ablation study for gene space mapping by excluding part of the strategy during the knowledge graph representation. In our approach, the relation-specific space mapping is based on a dynamic strategy \cite{TransD}, denoted as DynMap in the ablation study (the full version of our approach). An ablation version of our approach is replacing the relation-specific dynamic mapping of with a relation-specific static mapping \cite{TransR}, denoted as StaticMap in the ablation study. Since there is also relation-specific hyperplane strategy \cite{TransH}, and the relation-specific hyperplane can be regarded as a special case of relation-specific space, and the hyperplane projection is also a special case of mapping, therefore we also compare this relation-specific hyperplane strategy in the ablation study, denoted as PlaneMap. Moreover, we also include the representation without relation-specific consideration \cite{TransE} in the ablation study. In addition, the metric of distance in gene space representation is set to L1-norm in our approach, the metric can also be set to L2-norm in the ablation study. In the results of ablation study, the ROC curve demonstrates that for both STRING and iRefIndex interactions, the full version of our approach can yield better performance than those of the other cases (\textbf{Figure \ref{fig:ablationMeth}}). Also, our approach with sparsity-inducing L1-norm also outperforms that with Euclidean L2-norm (\textbf{Figure \ref{fig:ablationMeth}}). Generally, the full version of our approach with both dynamic mapping gene space and sparsity-inducing metric shows superiority to the other versions.

\subsection{Visualization for subtype-specificity}

\begin{figure}[h!]
	\begin{center}
		\includegraphics[width=0.7\linewidth]{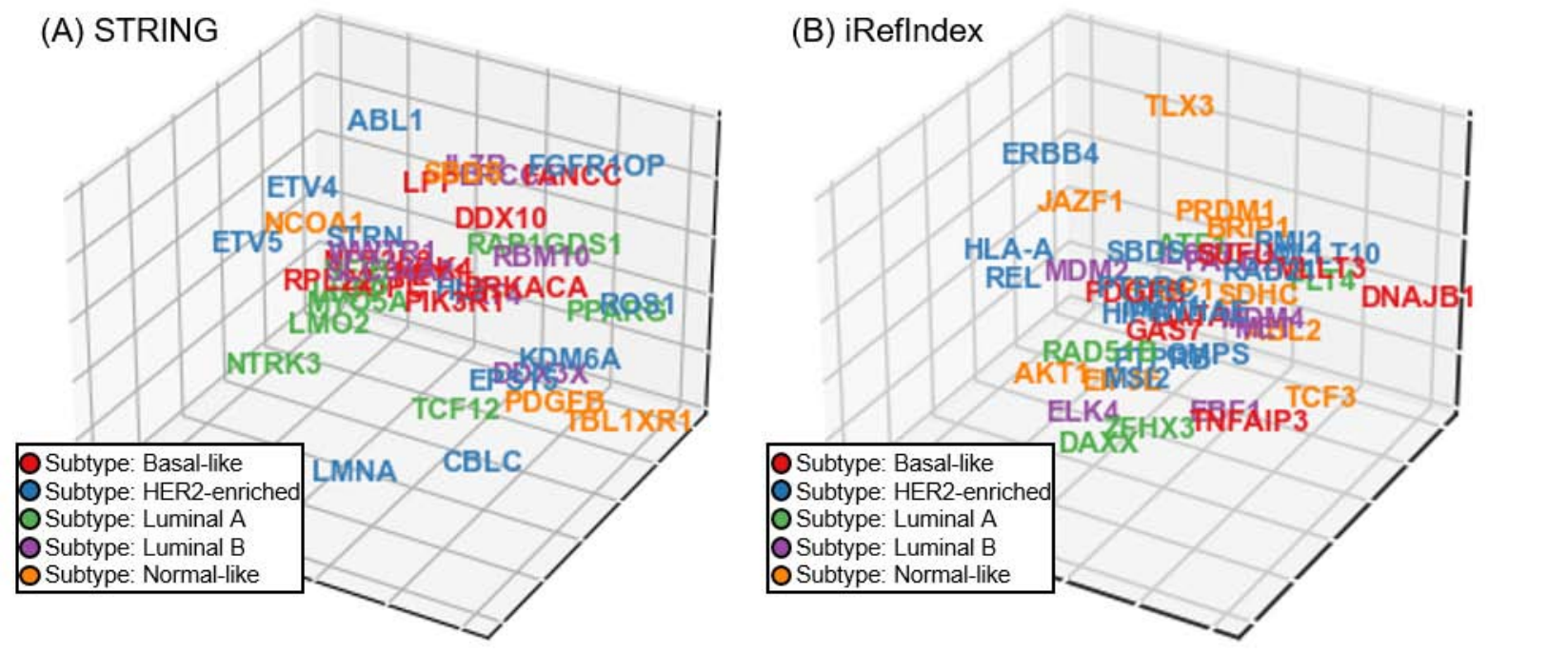}
	\end{center}
	\caption{The t-SNE 3D visualization for subtype-specificities of discovered drivers by our approach through. (A) Visualization result with interaction source of STRING. (B) Visualization result with interaction source of iRefIndex.}\label{fig:tsne}
\end{figure}

In addition to the aforementioned experiments on cancer driver discovery, subtype-specificity of the investigated driver is also another important aspect in precision medicine of cancer diagnosis. Unfortunately, unlike the cancer driver genes, subtype-specificity of genes do not have benchmarking. Accordingly, we perform a visualization experiment of subtype-specificity indication of our approach as an intuitive and visualized way \cite{MyBio}. In our approach, we can adopt the recovery loss score on the triplet (gene $G$, belong to $B$, subtype $S$) and obtain the loss score for subtype $S$. By switching the triplet tail across all subtypes in the triplet, we can compute the scores corresponding to the investigated gene across all the subtypes. Finally, we selected the subtype with minimizing recovery loss as the subtype-specificity of the gene to be tested. For discovered driver genes of our approach in the five-fold cross validation, we randomly select a fold and display the subtype-specificities of these genes. Since the dimension of gene space is usually larger than 3 and thus is difficult to display, we utilize the t-distributed neighbor embedding (t-SNE) on our gene space representation \cite{tSNE, tSNEGene}, and reduce the space dimension to 3D (\textbf{Figure \ref{fig:tsne}}). In the visualization of genes in the 3D space, we illustrate the distribution trends of different breast cancer subtypes through different colors. As shown in \textbf{Figure \ref{fig:tsne}A} and \textbf{\ref{fig:ablationData}B}, for the situations of both STRING and iRefIndex, we can observe distinct trends of each subtype (each color) across the space.

\subsection{Enrichment analysis for unscrambling functions}

\begin{table}
	\caption{The functional enrichment analysis results of the discovered drivers of our approach with interaction sources of STRING and iRefIndex respectively.}
	\begin{tabular}[t]{l | c | c | c }
		\hline
		Term (Results of STRING) & Percentage & p-value & FDR\\
		\hline
		hsa05200:Pathways in cancer & 27.90697674 & 2.10E-13 & 2.18E-11\\
		hsa05210:Colorectal cancer & 13.95348837 & 1.14E-11 & 5.95E-10\\
		hsa05215:Prostate cancer & 13.95348837 & 4.39E-11 & 1.52E-09\\
		hsa05205:Proteoglycans in cancer & 16.27906977 & 1.00E-09 & 2.60E-08\\
		hsa05213:Endometrial cancer & 10.46511628 & 4.97E-09 & 1.03E-07\\
		hsa05225:Hepatocellular carcinoma & 13.95348837 & 1.64E-08 & 2.85E-07\\
		hsa05220:Chronic myeloid leukemia & 10.46511628 & 4.44E-08 & 6.59E-07\\
		hsa01521:EGFR tyrosine kinase inhibitor resistance & 10.46511628 & 6.04E-08 & 7.26E-07\\
		hsa05226:Gastric cancer & 12.79069767 & 6.28E-08 & 7.26E-07\\
		hsa05216:Thyroid cancer & 8.139534884 & 1.74E-07 & 1.81E-06\\
		hsa05163:Human cytomegalovirus infection & 13.95348837 & 3.30E-07 & 3.02E-06\\
		hsa05221:Acute myeloid leukemia & 9.302325581 & 3.48E-07 & 3.02E-06\\
		hsa05218:Melanoma & 9.302325581 & 5.74E-07 & 4.26E-06\\
		hsa05223:Non-small cell lung cancer & 9.302325581 & 5.74E-07 & 4.26E-06\\
		hsa05224:Breast cancer & 11.62790698 & 6.71E-07 & 4.61E-06\\
		hsa05202:Transcriptional misregulation in cancer & 12.79069767 & 7.10E-07 & 4.61E-06\\
		hsa05214:Glioma & 9.302325581 & 7.61E-07 & 4.65E-06\\
		hsa05212:Pancreatic cancer & 9.302325581 & 8.33E-07 & 4.82E-06\\
		hsa05206:MicroRNAs in cancer & 15.11627907 & 1.11E-06 & 6.05E-06\\
		\hline
		\hline
		Term (Results of iRefIndex) & Percentage & p-value & FDR\\
		\hline
		hsa05200:Pathways in cancer & 23.65591398 & 5.59E-12 & 7.38E-10\\
		hsa05215:Prostate cancer & 11.82795699 & 5.44E-10 & 3.59E-08\\
		hsa05220:Chronic myeloid leukemia & 9.677419355 & 2.86E-08 & 9.43E-07\\
		hsa05212:Pancreatic cancer & 9.677419355 & 2.86E-08 & 9.43E-07\\
		hsa05225:Hepatocellular carcinoma & 11.82795699 & 1.14E-07 & 3.01E-06\\
		hsa04068:FoxO signaling pathway & 10.75268817 & 1.55E-07 & 3.40E-06\\
		hsa05223:Non-small cell lung cancer & 8.602150538 & 3.93E-07 & 7.40E-06\\
		hsa04630:JAK-STAT signaling pathway & 10.75268817 & 9.45E-07 & 1.56E-05\\
		hsa05210:Colorectal cancer & 8.602150538 & 1.33E-06 & 1.82E-05\\
		hsa04110:Cell cycle & 9.677419355 & 1.48E-06 & 1.82E-05\\
		hsa05166:Human T-cell leukemia virus 1 infection & 11.82795699 & 1.52E-06 & 1.82E-05\\
		hsa05224:Breast cancer & 9.677419355 & 4.72E-06 & 5.19E-05\\
		hsa05226:Gastric cancer & 9.677419355 & 5.21E-06 & 5.30E-05\\
		hsa05218:Melanoma & 7.52688172 & 7.04E-06 & 6.64E-05\\
		hsa05214:Glioma & 7.52688172 & 8.94E-06 & 7.87E-05\\
		hsa05161:Hepatitis B & 9.677419355 & 9.68E-06 & 7.99E-05\\
		hsa04919:Thyroid hormone signaling pathway & 8.602150538 & 1.31E-05 & 1.02E-04\\
		hsa05206:MicroRNAs in cancer & 11.82795699 & 2.92E-05 & 2.14E-04\\
		hsa05202:Transcriptional misregulation in cancer & 9.677419355 & 3.46E-05 & 2.37E-04\\
		\hline
	\end{tabular}
\end{table}

Although the aforementioned experiments have demonstrated the results of both cancer driver discovery and subtype-specificity indication of the discovered potential driver genes, the biological functions of these genes are still not unscrambled. Hence, we further apply the functional enrichment analysis on our discovered potential driver genes to find which biological functions are related to these genes. Here we adopt the functional enrichment analysis through a widely-used platform called Database for Annotation, Visualization and Integrated Discovery (DAVID) \cite{DAVID}. Since the top scored genes are expected to have more potential of being cancer drivers, we select the top 100 discovered genes into the functional enrichment analysis. The functional enrichment analysis yields the enriched functional terms of curated by Kyoto Encyclopedia of Genes and Genomes (KEGG) \cite{KEGG}. Table 1 shows the top twenty functional terms unscrambled by the enrichment analysis. Through the unscrambling results, we can find that the top enriched functional terms are highly associated with cancer process, such as pathways in cancer, EGFR tyrosine kinase inhibitor resistance, JAK-STAT signaling pathway, cell cycle, microRNAs in cancer, transcriptional misregulation in cancer, and breast cancer. Generally speaking, by enrichment analysis Unscrambling, the discover potential driver genes of our approach are observed to be functionally associated with cancer process.

\section{Discussion}

Discovering subtype-specific drivers is an inevitable demand in breast cancer precision medicine. The existing widely-used computational tools for driver discovery mainly focus on exploiting the information from DNA aberrations or gene interactions. However, recent studies have demonstrated that expect DNA aberrations, RNA alternations also play important roles in cancer driver events \cite{NatureRNA}, but there still lacks of an integration strategy for multiple types of aberrations from both DNA and RNA. Generally, there are mainly two reasons make the integration as a challenging task. The first reason is the data format incompatibility, where the data formats of different types of aberrations are distinct from each other. The second reason is the aberration type heterogeneity, where the patterns of multi-type aberrations also vary from each other. In this paper, we propose an integration strategy for subtype-specific breast cancer driver discovery with a "splicing-and-fusing" framework. In our framework, to address the data format incompatibility, the "splicing-step" employs knowledge graph structure to connect multi-type aberrations from the DNA and RNA data into a unified formation \cite{ProIEEE}. To tackle the aberration type heterogeneity, the "fusing-step" adopts a dynamic mapping gene space integration approach to represent the multi-type information by vectorized profiles \cite{ReviewKG}. The evaluation experiments also demonstrate the advantages of our approach in both the integration of multi-type aberrations from DNA and RNA and the discovery of subtype-specific breast cancer drivers.

The main advantages of our study can be summarized as three points. The first advantage is that the multi-type aberrations can contain more information beyond the scope of breast cancer genomic data. By adding the RNA alternations, interactions between genes, the subtype annotations of samples, and the supervision information of known benchmarking drivers, the discovery task of subtype-specific cancer drivers has a richer connotation than those of most of the existing computational tools. The second advantage is that the knowledge graph structure can successfully overcome the issue of data format incompatibility. We can redescribe the existing data with a series of triplets of facts, and splice them together by connecting common entities and relations into an integrated knowledge graph of multi-type aberrations of breast cancer. The third advantage is that the idea of dynamic mapping strategy can fuse the triplet data into relation-specific gene space, and the low-dimensional fused gene space can describe the aberration type heterogeneity as many-sided representations. Through the advantages above, our approach of multi-type aberration integration can not only discover potential driver genes from breast cancer data, but also indicate the subtype-specificities of these potential drivers by the integration of both DNA and RNA data of breast cancer samples.

Despite the achievements of our integrated approach, there are also some limitations in our study. An inevitable limitation is that our approach relies on a strong restriction of the alignment of the DNA and RNA data samples. If some breast cancer samples only contain DNA data or RNA data, these samples cannot be used in our integration analysis \cite{MultiOmic}. Actually, there are less than one thousand cancer samples that satisfy the requirement of combination of both DNA and RNA data in the TCGA database used in our study. Also, since the gene space representations are expected to be information preserving, the learning process of these representations requires more computational resources than those of the traditional statistical based methods. As for the future plan of this study, a promising future work of our study is to expand the framework to the pan-cancer analysis \cite{HotNet2}. In consideration of the capability of the multi-type aberration fusion and the subtype-specificity indication, the integration of various cancer types in pan-cancer analysis is also feasible in some extent. In summary, we design a "splicing-and-fusing" framework with knowledge graph connection and dynamic mapping gene space fusion of multi-type aberrations data from DNA and RNA, and the application of our approach on breast cancer samples can successfully discover potential cancer drivers with subtype-specificity indication.

\section*{Conflict of Interest Statement}
The authors declare that the research was conducted in the absence of any commercial or financial relationships that could be construed as a potential conflict of interest.

\section*{Author Contributions}
JX, ZD and WS contributed to the conception and design of the study. JX, YL and WS designed and implemented the experiments of data integration.
JX, ZD and WQ performed data analysis for the study. 
JX, YL, and QW prepared the original manuscript, figures and tables. JX and WS reviewed and revised the manuscript.

\section*{Funding}
This work is supported by the National Natural Science Foundation of China (Grant Nos. 61901322, 62202117, and 61974109). This work is also supported by the Special Foundation in Department of Higher Education of Guangdong (Grant No. 2022ZDX2053).
The funding body did not play any roles in the design of the study and collection, analysis, and interpretation of data and in writing the manuscript.

\section*{Acknowledgments}
We would like to thank Dr Liying Yang for her felpful suggestions.

\bibliographystyle{unsrt}

\end{document}